# Conduction in molecular junctions: Inelastic effects

Dvira Segal and Abraham Nitzan

School of Chemistry, The Sackler Faculty of Science, Tel Aviv University,
Tel Aviv, 69978, Israel

18/10/2001 13:03

## Abstract

# 1. Introduction

The investigation of electrical junctions in which single molecules or small molecular assemblies operate as conductors connecting 'traditional' electrical components such as metal or semiconductor contacts constitutes a major part of what has become the active field of molecular electronics. Their diversity and versatility and amenability to control and manipulation make molecules and molecular assembly potentially important components in nano-electronic devices. Indeed basic properties such as single electron transistor behavior and current rectification have already been demonstrated. At the same time major difficulties lie on the way to real technological applications. These difficulties stem on one hand from problems associated with the need to construct, characterize, control and manipulate small molecular structures at confined interfaces with a high degree of reliability and reproducibility. On the other hand lie issues of stability of such small junctions. One cause for concern is heat generation and dissipation in these systems.[1,2] In this paper we discuss this issue using simple models for molecular bridges connecting two metal contacts.

It has long been recognized that tunneling electrons interact, and may exchange energy, with nuclear degrees of freedom in the tunneling medium. One realization of such processes is inelastic electron tunneling spectroscopy.[3] Inelastic electron tunneling may also cause chemical bond breaking and chemical rearrangement in the tunneling medium, either by electron induced consecutive excitation or via transient formation of a negative ion.[4-7]-[1]. The phenomenology of inelastic electron transmission is also closely related to other electronic processes in which transient occupation of an intermediate state drives a phonon field. [8] Intramolecular vibrational excitation in resonant electron scattering,[9] phonon excitation in resonant electron tunneling in quantum-well heterostructures[10] and electron induced desorption[11,12] can all be described on within this framework.

Nuclear motion, associated with solvent reorganization about the donor and acceptor sites upon the change in their charge state, is an essential ingredient of the standard theory of electron transfer. When the donor and acceptor are replaced by metal contacts this aspect of the process is not crucial anymore, because the metal environments can supply and drain charge carriers from the system without nuclear rearrangement. It is the issue of how thermal relaxation *in the molecular bridge* connecting the metals affects the transport process that becomes central. It should be emphasized that this is an important, even if usually overlooked, issue also in 'regular' electron transfer processes, see, e.g. [13].

The Medvedev-Stuchebrukhov theory[14] corresponds to the lowest order correction due to intermediate state nuclear relaxation for the electron transfer rate in the so called superexchange processes where the electron transfer is mediated by intermediate (bridge) high energy electronic states. On the other extreme side we find sequential processes that are best described by two or more consecutive electronic transitions. Obviously, intermediate situations can exist, see e.g., [15-19]

Closely related to this phenomenology is the process of light scattering from molecular systems where the donor and acceptor states are replaced by the incoming and outgoing photons. Elastic (Rayleigh) scattering is the analog of the 2-state 'standard' electron transfer process. Inelastic (Raman) scattering and resonance Raman scattering involve intermediate states coherently. Resonance fluorescence is the process that takes place after thermal relaxation and dephasing occured in the intermediate state manifold.

The importance of dephasing effects in the operation of microscopic junctions has long been recognized. [20,21] Most of the work on dephasing effects in mesoscopic solid-state junctions follows the work of Büttiker[22] who has introduced phase destruction processes by conceptually attaching an electron reservoir onto the constriction connecting the metal contacts. A different origin of dephasing is implied by the random coupling model for long-range electron transfer of Bixon and Jortner[23,24]. Recent applications of hopping models for electron transfer processes in DNA[25-27] assume that dephasing predominates these processes. While coupling to the thermal environment is implicit in these treatments, several groups have recently discussed models for electron transport with explicit coupling to phonons [28-34]. These works provide exact numerical solutions of simple models: 1-dimensional tight binding transport model, only a few harmonic oscillators and essentially zero temperature systems. An alternative approach uses the machinery of non-equilibrium statistical mechanics, starting from a Hamiltonian for the system, bath and system-bath interaction and projecting out the bath degrees of freedom. [35-43]. The resulting reduced equations of motion for the electronic subsystem contain, in addition to the deterministic part that describes transport in the isolated system, also dephasing and energy relaxation rates that are related explicitly to properties of the thermal and the system-bath coupling. A recent development of this approach[44] makes it possible to examine the final energy distribution of the transmitted electron for a given incident energy. Such final energy resolution is not an observable in this kind of electron transmission experiments (in contrast to, e.g. photoemission where the final state of the emitted electron is directly observable), however this information is needed for evaluating the current at finite voltage drop across the junction, and can be also used to evaluate the heat 'left behind' in the bridge, an important element in estimating heating effects in electron transmission through molecular junctions.[45]

These works, that lead to reduced equation of motion for the density operator in the electron-molecule subsystem, treat the molecule-thermal bath interaction in the weak-coupling limit, using variants of the Redfield approximation[46,47] that describes this coupling using second order perturbation theory. The other limit, with full thermal relaxation assumed at any bridge site ocupied by the electron has been anticipated in classical hoping models that were recently used for, e.g., DNA-bridge mediated electron transfer.[25-27,48] A general treatment that can in principle reduce to these two limits have been presented only for the three-level system (including the donor and acceptor levels), i.e. where the bridge involves only one intermediate electronic state.[17,18] or, using a path integral approach, for thermal relaxation effects in tunneling through continuous potential barriers.[49,50], see also [51])

The present paper supplements our recent study of thermal relaxation effects in steady state electron transmission through a molecular bridge. Section 2 reintroduces the model: A tight binding model for the bridge, with coupling on the left and on the right to continua representing metal electrodes. For the weak coupling limit we present in Section 3 an analysis that improves the calculation of Ref. [44] of the final energy distribution of the transmitted electron. In section 4 we analyze the strong coupling limit using an approach that combines the small polaron transformation[52-54] with the Redfield approximation[46,47]. Section 5 analyzes our findings in the light of what is known on the behavior of the minimal 3-state model. Section 6 concludes.

## 2. Model and notations

At the focus of our consideration is a molecular bridge (M) described by a tight binding model with $N$ sites associated with a set of states $\{n\}$, one per site, taken for simplicity to be mutually orthogonal with nearest-neighbor coupling. The corresponding Hamiltonian is

$$H_M = H_0 + V$$

$$H_0 = \sum_{n=1}^{N} E_n \mid n \!>\! <\! n \mid \quad ; \quad V = \sum_{n=1}^{N-1} V_{n,n+1} \mid n \!>\! <\! n+1 \mid + V_{n+1,n} \mid n+1 \!>\! <\! n \mid \tag{1}$$

This bridge connects two metal electrodes, represented within the model by continuous manifolds of non-overlapping states: the 'left' continuum ($L$), and the 'right' continuum ($R$), which are assumed to couple only to the first (1) and last ($N$) bridge states, respectively. In what follows these continuous manifolds are sometimes denoted collectively by $J$, i.e., $J=(L,R)$. These manifolds are characterized by their density of states, $\rho_L(E)$ and $\rho_R(E)$. The corresponding Hamiltonians and couplings are written in the forms

$$H_J = \sum_l E_l \mid l \!>\! <\! l \mid + \sum_r E_r \mid r \!>\! <\! r \mid \tag{2}$$

$$H_{JM} = \sum_l V_l + \sum_r V_r$$

$$V_l = V_{l,1} \mid l \!>\! <\! 1 \mid + V_{1,l} \mid 1 \!>\! <\! l \mid \qquad V_r = V_{r,N} \mid r \!>\! <\! N \mid + V_{N,r} \mid N \!>\! <\! r \mid \tag{3}$$

Here $\{l\}$ and $\{r\}$ denote the left and right continuum states, respectively. Fig. 1 displays a schematic diagram of this system.

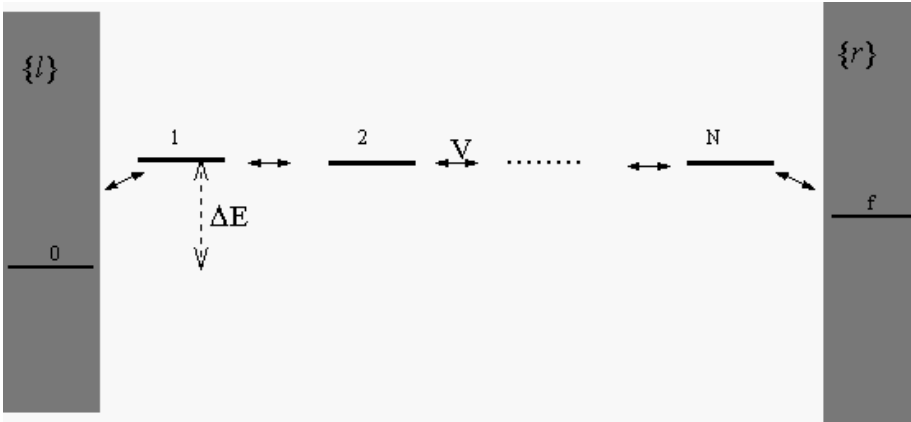

*Fig. 1.* A schematic diagram of the model used to discuss thermal relaxation effects in conduction through molecular bridges. See text for details.

Finally, the thermal environment is represented by the 'bath' Hamiltonian $H_B$, and thermal relaxation is assumed to be effective only in the molecular subspace. The molecule-thermal bath coupling is taken to be of the form

$$F = \sum_{n=1}^{N} F_{n,n} \mid n \!>\! <\! n \mid \tag{4}$$

Where $F$ are operators in the bath subspace. The exact form of $F$ is not important, but in the present discussion, we will assume that the coupling to the thermal environment is weak. The coupling term is characterized by its correlation function, whose Fourier transforms satisfy the detailed balance relation

$$\int\limits_{-\infty}^{\infty} dt e^{i\omega t} < F_{n,n}(t)F_{n,n}(0) > = e^{\beta\hbar\omega} \int\limits_{-\infty}^{\infty} dt e^{i\omega t} < F_{n,n}(0)F_{n,n}(t) > \qquad ; \qquad \beta = \left(k_B T\right)^{-1} (5)$$

where $T$ is the temperature and $k_B$ - the Boltzmann constant. For specificity we will sometime use

$$< F_{n,n}(t)F_{n',n'}(0) > = \delta_{n,n'} \frac{\kappa}{2\tau_c} \exp\left(-|t|/\tau_c\right) \qquad\qquad (6)$$

in which $\kappa$ and $\tau_c$ play the roles of coupling strength and correlation time, respectively. The RHS of Eq. (6) becomes $\kappa\delta_{n,n}\delta(t)$ in the Markovian, $\tau_c \to 0$, limit.

The Hamiltonian of the overall system (molecular bridge, electrodes, thermal environment and the corresponding couplings is

$$H = H_M + H_B + F + H_J + H_{JM} \qquad\qquad (7)$$

We consider a steady state pumped by a particular incoming state |0> of the L manifold. In the absence of thermal interactions the time evolution of the density matrix may be obtained from the amplitude equations

$$\dot{c}_0 = -iE_0 c_0 \qquad\qquad (8)$$

$$\dot{c}_1 = -iE_1 c_1 - iV_{1,0}c_0 - iV_{1,2}c_2 - i\sum_{l\neq 0} V_{1,l}c_l \qquad\qquad (9)$$

$$\dot{c}_n = -iE_n c_n - iV_{n,n-1}c_{n-1} - iV_{n,n+1}c_{n+1} \quad ; \quad n = 2,...,N-1 \qquad (10)$$

$$\dot{c}_N = -iE_N c_N - iV_{N,N-1}c_{N-1} - i\sum_r V_{N,r}c_r \qquad\qquad (11)$$

$$\dot{c}_l = -iE_l c_l - iV_{l,1}c_1 - (\eta/2)c_l \qquad\qquad (12)$$

$$\dot{c}_r = -iE_r c_r - iV_{r,N}c_N - (\eta/2)c_r \qquad\qquad (13)$$

where in Eqs. (12) and (13) the rate $\eta$, which is taken to 0 at the end of the calculation, is a mathematical apparatus that insures the outgoing nature of the states in the L and R continua. At long time, this system approaches a steady state where the amplitude of each state oscillates according to $c_j = C_j e^{-iE_0 t}$ ; $j = \{n\},\{l\},\{r\}$ . The normalized steady state transmitted flux is $k_{0\to R} = \eta \sum_r |C_r|^2 / |C_0|^2$ and the transmission coefficient for initial and final energies $E_0$ and $E$ is $\mathcal{T}_{el}(E_0, E) = 2\pi\rho_L(E_0)k_{0\to R}$ . (The subscript "$el$" is used to denote the elastic character of this transmission process). This leads to[44]

$$\mathcal{T}_{el}(E_0, E) = \delta(E - E_0)\mathcal{T}_{el}(E_0) = \delta(E - E_0)Tr_N\left(G^M(E_0)\Gamma^L(E_0)G^M(E_0)\Gamma^R(E_0)\right)$$

$$(14)$$

(we use $\mathcal{T}'$ to denote the differential (per unit energy range) transmission coefficient, while $\mathcal{T}_{el}$(E) is the elastic transmission coefficient) where $G^{(M)}(E)$ is the Green's function associated with the subspace of the molecular bridge

$$G^{(M)}(E) = \left(E - \mathbf{H}^{(M)}(E)\right)^{-1}$$

$$(15)$$

$$H_{n,n'}^{(M)}(E) = E_n\delta_{n,n'} + V_{n,n'} + \Sigma_{n,n'}(E)$$

$$(16)$$

and where $\Sigma$ is the self energy associated with the interaction of the bridge states with the metal electrodes and $\Gamma$ – its imaginary part

$$\Sigma_{n,n'}(E) = \Sigma_{n,n'}^{(L)}(E) + \Sigma_{n,n'}^{(R)}(E)$$

$$\Sigma_{n,n'}^{(J)}(E) = \sum_j \frac{V_{n,j}V_{j,n'}}{E - E_j + i\eta/2} = \Lambda_{n,n'}^{(J)}(E) - \frac{1}{2}i\Gamma_{n,n'}^{(J)}(E) \; ; \; J = L, R$$

$$(17)$$

The elastic transmission coefficient $\mathcal{T}_{el}$(E) is related to the zero bias conduction of the junction by the Landauer formula[21,55]

$$g = \frac{e^2}{\pi\hbar}\mathcal{T}_{el}(E_F)$$

$$(18)$$

where $E_F$ is the Fermi energy.

## 3. Thermal relaxation effects in the weak coupling limit

It is convenient, for notational simplicity to consider the case of one bridge level, $N$=1. Generalization of the procedure described below to many bridge levels is straightforward. For N=1 Eqs. (9)-(11) coalesce into

$$\dot{c}_1 = -iE_1c_1 - iV_{1,0}c_0 - i\sum_{l\neq 0}V_{1,l}c_l - i\sum_{r\neq 0}V_{1,r}c_r$$

$$(19)$$

In the presence of thermal interactions the system has to be described in terms of its density matrix. The time evolution equations for the density matrix elements $\rho_{ab} = c_a c_b^*$ are easily derived in the absence of thermal interactions from Eqs. (8)-(13). This should be supplemented by terms arising from the system-bath interaction. At steady state this leads to[44] (for $N$=1)

$$\rho_{00} = \text{constant}$$

$$(20)$$

$$\dot{\rho}_{0,1} = 0 = -iE_{0,1}\rho_{0,1} + iV_{0,1}\rho_{0,0} + i\sum_j V_{j,1}\rho_{0,j} - i[F,\rho]_{0,1} \qquad (21)$$

$$\dot{\rho}_{1,1} = 0 = -iV_{1,0}\rho_{0,1} + iV_{0,1}\rho_{1,0} - i\sum_j V_{1,j}\rho_{j,1} + i\sum_j V_{j,1}\rho_{1,j} - i[F,\rho]_{1,1} \qquad (22)$$

$$\dot{\rho}_{0,j} = 0 = -iE_{0,j}\rho_{0,j} + iV_{1,j}\rho_{0,1} - (\eta/2)\rho_{0,j} - i[F,\rho]_{0,j} \qquad (23)$$

$$\dot{\rho}_{1,j} = 0 = -iE_{1,j}\rho_{1,j} - iV_{1,0}\rho_{0,j} - i\sum_{j'}V_{1,j'}\rho_{j',j} + iV_{1,j}\rho_{1,1} - (\eta/2)\rho_{1,j} - i[F,\rho]_{1,j} \quad (24)$$

$$\dot{\rho}_{j',j} = 0 = -iE_{j',j}\rho_{j',j} - iV_{j',1}\rho_{1,j} + iV_{1,j}\rho_{j',1} - \eta\rho_{j',j} \qquad (25)$$

where the index $j$ corresponds to states from both the left and the right manifolds and where $E_{a,b} \equiv E_a - E_b$. The matrix elements $\rho_{a,b}$ are now operators in the bath space. It should be noted that Eqs. (20)-(25) deviate in important ways from the standard form obtained from the Liouville equation $d\rho/dt = -i[H,\rho]$.[44]

In Ref. [44] we have described a procedure to evaluate the energy resolved steady state flux in this system. This procedure was based on the simplifying assumption that the terms involving the thermal interaction $F$ could be disregarded in Eqs. (23)-(25) that involve the continuous manifolds J={$j$}. Under this assumption the effect of these manifolds on the time evolution of Eqs. (21) and (22) can be represented by appropriate self energy elements as in Eq. (16), so that Eqs. (21)-(25) become

$$\dot{\rho}_{0,1} = 0 = -i\left(E_0 - \tilde{E}_1\right)\rho_{0,1} + iV_{0,1}\rho_{0,0} - \frac{1}{2}\Gamma_1\rho_{0,1} - i[F,\rho]_{0,1} \qquad (26)$$

$$\dot{\rho}_{1,1} = 0 = -iV_{1,0}\rho_{0,1} + iV_{0,1}\rho_{1,0} - \Gamma_1\rho_{11} - i[F,\rho]_{1,1} \qquad (27)$$

$$\dot{\rho}_{0,j} = 0 = -i(E_0 - E_j)\rho_{0,j} + iV_{1,j}\rho_{0,1} - (\eta/2)\rho_{0,j} \qquad (28)$$

$$\dot{\rho}_{1,j} = 0 = -i(\tilde{E}_1 - E_j)\rho_{1,j} - iV_{1,0}\rho_{0,j} + iV_{1,j}\rho_{1,1} - \frac{1}{2}\Gamma_1\rho_{1,j} \qquad (29)$$

$$\dot{\rho}_{j',j} = 0 = -i(E_{j'} - E_j)\rho_{j',j} - iV_{j',1}\rho_{1,j} + iV_{1,j}\rho_{j',1} - \eta\rho_{j',j} \qquad (30)$$

where $\Gamma_1 = \Gamma_{11}(E_0)$ and where $\tilde{E}_1 = E_1 + \Lambda_{11}(E_0)$. $\Gamma_{11}$ and $\Lambda_{11}$ were defined in Eq. (17).

Our ultimate goal is to obtain the evolution of the reduced system's density matrix $\sigma = Tr_B\rho$. This trace can be done trivially in Eqs. (28)-(30) that do not contain the heat bath. At the same time Eqs. (26) and (27) that, together with the boundary condition $\rho_{00}$=constant, describe a steady state in a damped and thermally relaxing two-level system, can be handled independently from Eqs. (28)-(30). In the weak system-thermal bath coupling this is done using the Redfield approximation, yielding steady states expressions for the reduced density matrix elements $\sigma_{11}$, $\sigma_{10}$ and $\sigma_{01}$. Using these in the reduced forms of Eqs. (28)-(30) yields the desired energy resolved transmission as described in Ref. [44]. For example, for the case of a single bridge level considered here, and in the strongly off-resonance case where the energy gap $E_1$-$E_0$ is much larger than all other energy parameters in the system, i.e., $E_1 - E_0 \gg |V_{10}|, \kappa, \Gamma_1$ ($\kappa$ is defined in Eq. (6), and we are considering the Markovian limit, $\tau_c$=0) the differential (final energy resolved) transmission coefficient is obtained in the form [44]

$$\mathcal{T}'(E_0, E) = \mathcal{T}_{el}(E_0)\left[\delta(E_0 - E) + \frac{(\kappa/2\pi)e^{-\beta(E_1 - E_0)}}{(E_1 - E)^2 + (\Gamma_1/2)^2}\right] \qquad (31)$$

where, again, $\mathcal{T}_{el}(E_0)$ is the elastic transmission coefficient. The final-energy resolved transmission is seen to consist of an elastic contribution supplemented by an inelastic, thermally activated terms.[1] However a shortcoming of the approximation used here is seen in the fact that the elastic contribution appears to be independent of the coupling to the thermal environment. In fact we know that this contribution, the analog of the zero-phonon peak in optical and Raman spectroscopy of molecules embedded in condensed environments, does contain thermal effects. While in our present application the expected corrections are small and probably negligible, in the broader context of quantum transport theory it is of interest to consider improvements on the approximation used above.

Such an improvement can be achieved by realizing that the essence of the approach that lead to Eqs. (8)-(13) and (20)-(25) is to use the incoming state |0> as a driving term in the steady state dynamics. In the context of system/bath models this is a state that belongs to one of the baths (e.g. the left metal lead) that, because of its special role as an incoming state, is left in the system's subspace in the reduction process that leads to equations of motion for system's variables. A generalization of this approach is obtained by including also the final (outgoing) state of the scattering process under consideration, a state of the accepting continuum (e.g. the metal lead on the right) in the system's subspace, again because of its special role as the state that is finally detected.[2] Appendix A illustrates both ways of reducing the description of the system's dynamics for the case of a single bridge level without thermal relaxation effects. In this case both procedures are shown to yield the same result for the transmission, and the one that handles the initial and final states symmetrically does not have any advantage over the less symmetrical way taken before. However, this new approach provides a better route in the presence of thermal interactions as we now show.

---

[1] These contributions are separable only in the limit described above Eq. (31).

Again we limit ourselves for simplicity to a model with only a single bridge level. We start from a set of equations similar to (26)-(30) but written so as to treat the incoming and outgoing states symmetrically.

$$\rho_{0,0} = \text{constant} \tag{32}$$

$$\dot{\rho}_{0,1} = 0 = -i\left(E_0 - \tilde{E}_1\right)\rho_{0,1} + iV_{0,1}\rho_{0,0} - \frac{1}{2}\Gamma_1\rho_{0,1} - i[F,\rho]_{0,1} \tag{33}$$

$$\dot{\rho}_{1,1} = 0 = -iV_{1,0}\rho_{0,1} + iV_{0,1}\rho_{1,0} - iV_{1,f}\rho_{f,1} + iV_{f,1}\rho_{1,f} - \Gamma_1\rho_{11} - i[F,\rho]_{1,1} \tag{34}$$

$$\dot{\rho}_{0,f} = 0 = -i(E_0 - E_f)\rho_{0,f} + iV_{1,f}\rho_{0,1} - (\eta/2)\rho_{0,f} \tag{35}$$

$$\dot{\rho}_{1,f} = 0 = -i(\tilde{E}_1 - E_f)\rho_{1,f} - iV_{1,0}\rho_{0,f} + iV_{1,f}\rho_{1,1} - iV_{1,f}\rho_{f,f} - \frac{1}{2}\Gamma_1\rho_{1,f} - i[F,\rho]_{1,f} \tag{36}$$

$$\dot{\rho}_{f,f} = 0 = -iV_{f,1}\rho_{1,f} + iV_{1,f}\rho_{f,1} - \eta\rho_{f,f} \tag{37}$$

These equations describe scattering from the incoming state 0 to the final outgoing state *f*. All other states in the continuous manifolds {*r*} and {*l*} were projected out and consequently the bridge state |1> acquires a shift and damping terms. As discussed in Appendix A, in the large system limit (i.e. when the manifolds L and R become true continua) these shift and damping are the same as taken in Eqs. (26)-(30)). However, in contrast to Eqs. (26)-(30) the coupling to the thermal environment is not disregarded in Eq. (36).

Next, assuming weak system-thermal bath coupling, the Redfield approximation is invoked to reduce Eqs. (32)-(37) into equations for the density matrix elements $\sigma_{i,j}$ in the 'system's ' subspace (*i*, *j* = 0,1,*f*) as outlined in Appendix B. The resulting equations

---

for $\sigma_{0,1}$, $\sigma_{1,0}$, $\sigma_{1,1}$, $\sigma_{0,f}$, $\sigma_{f,0}$, $\sigma_{1,f}$, $\sigma_{f,1}$, $\sigma_{f,f}$ yield $\sigma_{f,f}$, and consequently the final-energy resolved flux $\eta\sigma_{f,f}$.

Even for this simple case, and certainly in the case of a general bridge where the state |1> is replaced by a set of bridge states {|n>}, the resulting expression for the transmission is two cumbersome to display and discuss analytically. Instead we show some numerical results that compare the result of the present computational scheme with that employed earlier.[44] Two general observations are of interest:

(a) On the technical side, our present approach employs different reduction schemes for different final states. This arises from the fact that the 'system' associated with the effective Hamiltonian, Eq. (103), that needs to be diagonalized in the Redfield scheme depends on the particular final state observed. Consequently, the steady state sum rule $\eta\sum_r \sigma_{r,r} = \Gamma_1^{(R)}\sigma_{1,1}$ that states that the differential flux integrated over the $R$ manifold is equal to the rate at which the bridge supplies population into that manifold, provides a non-trivial consistency check on this procedure. We find that this sum rule is indeed satisfied in the large system limit ($\Omega \to \infty$), where $V_{1,r} \to 0$ while $\rho_R \to \infty$ so that $\Gamma_1^{(R)}(E) = 2\pi\sum_r |V_{1,r}|^2 \delta(E - E_r)$ is finite.

(b) Taking care to consider the proper scattering ($\Omega \to \infty$) limit, our earlier calculation[44] provides a reasonable approximation for the energy resolved transmission. In particular, when the energy and coupling parameters allow the separation of the tunneling and activated fluxes, the activated component obtained in the present approach is practically the same within the numerical accuracy of our calculation as that obtained before. The consequence of this observation is that the conclusions of our earlier work[42-44] concerning the temperature and bridge-length dependence of the transmission, in particular the prediction of transition from tunneling to activated transmission at increasing temperature and bridge length remain intact. As an example, Fig. 2 shows the transmission probability at $T$=300K as a function of bridge length $N$ in the model of Fig. 1, using the model parameters $\Delta E$=3000cm$^{-1}$, $V \equiv V_{l,1} = V_{1,r}$=200cm$^{-1}$ (same for all levels of the L and R manifolds), $\kappa$=10cm$^{-1}$ (the thermal bath is assumed Markovian, i.e. $\tau_c$=0) and $\Gamma^{(L)} = \Gamma^{(R)}$=160cm$^{-1}$. The transition from an exponential dependence on $N$ to transmission that is practically $N$ independent[3] marks the onset of the activated transmission. This behavior has been recently observed in electron transfer through DNA bridges.[56]

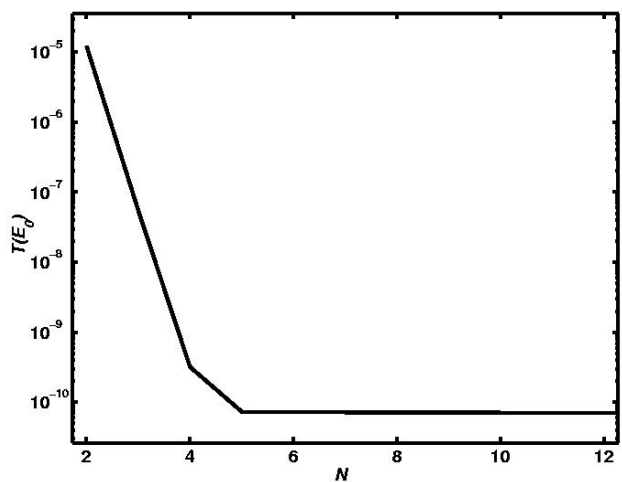

Fig. 2. The transmission coefficient as a function of number of bridge sites calculated for the model of Fig. 1 with $\Delta E$=3000cm$^{-1}$ $\Gamma^{(L)}$=$\Gamma^{(R)}$=160cm$^{-1}$, $V$=200 cm$^{-1}$, $T$=300K, $\kappa$=10 cm$^{-1}$. ($\eta$=0.1 was used in this and the following calculation, however the result does not depend on this particular choice).

---

[3] (as shown in Refs. (42) D Segal, A Nitzan, WB Davis, MR Wasilewski, MA Ratner: Electron transfer rates in bridged molecular systems 2: A steady state analysis of coherent tunneling and thermal transitions. J. Phys. Chem. B 104 (2000) 3817.

(43) D Segal, A Nitzan, MA Ratner, WB Davis: Activated Conduction in Microscopic Molecular Junctions. J. Phys. Chem. 104 (2000) 2790. the transmission depends on N as $(\alpha_1 + \alpha_2 N)^{-1}$, where often $\alpha_1 \gg \alpha_2$)

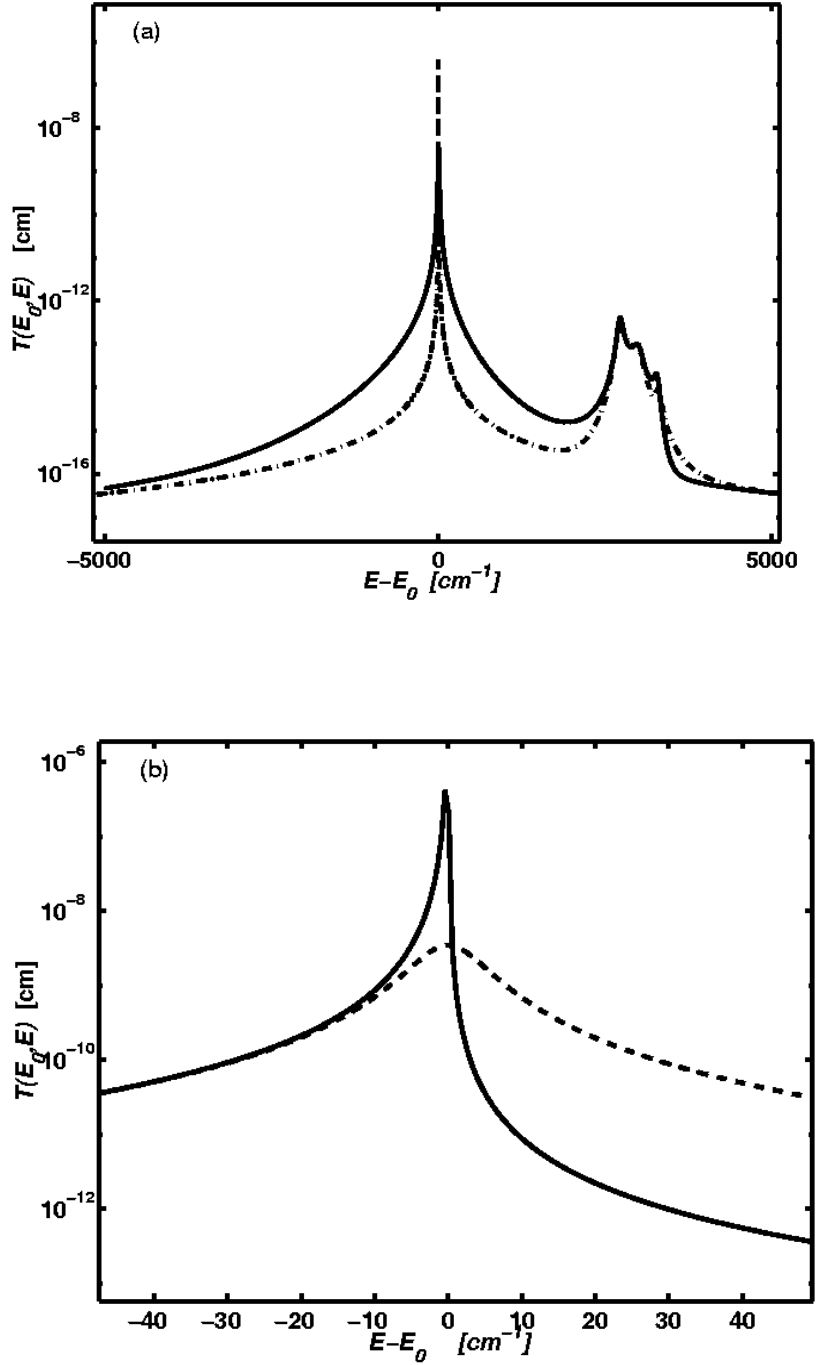

Fig. 3. The energy resolved transmission calculated for the model of Fig. 1 with $\Delta E$=3000cm$^{-1}$, $\Gamma^{(L)}$=$\Gamma^{(R)}$=160cm$^{-1}$, $V$=200 cm$^{-1}$ $\kappa$=10 cm$^{-1}$, $N$=3. (a) $T$=300K. The full line is the result of the present theory. The dashed line results from the theory of Ref. [44]. (b). A closeup on the quasi-elastic peak computed using the present theory for the same model parameters. Full line - $T$=0K. Dashed line - $T$=300K.

Fig. 3 shows the energy resolved transmitted flux obtained from this calculation for the model of Fig. 1 characterized by the parameters: $N$=3, $\Delta E$=3000cm$^{-1}$, $V$=200cm$^{-1}$ $\kappa$=10cm$^{-1}$ (the thermal bath is assumed Markovian, i.e. $\tau_c$=0) and $\Gamma^{(L)} = \Gamma^{(R)}$=160cm$^{-1}$.

For obvious technical reasons the width parameter $\eta$ cannot be taken zero in the numerical calculation, and $\eta$=0.1cm$^{-1}$ is used here. Fig. 3a compares the results obtained for temperature $T$=300K using the method described above (full line) and the earlier approach of Ref. [44] (dashed line). Energy is measured relative to that of the incident state. The thermally activated component at about $E$=3000cm$^{-1}$ shows three peaks (corresponding to the three bridge levels) and is practically the same in the two calculations. The difference between the present and the earlier approaches affects mainly the quasi-elastic transmission component about $E$=0. While the earlier approach yields a temperature-independent elastic peak of zero width (the width of the dashed line in Fig. 3 results from using a finite $\eta$ in the numerical calculation), we now have a quasi-elastic peak of finite width that shows a characteristic asymmetry about $E$=0. This is seen in Fig 3b which shows the energy resolved transmission in the neighborhood of the quasi-elastic peak at $T$=0K (full line) and $T$=300K (dashed line). The temperature dependence of the quasi-elastic peak is shown in more detail in Fig. 4. The peak becomes increasingly asymmetric as the temperature decreases and at the same times shifts to lower energies; both effects resulting from the increasing domination of phonon emission processes. Fig. 5 depicts $\mathcal{T}(E)$/ $\mathcal{T}(-E)$ as a function of the final energy $E$, measured relative to the energy of the incident state, showing that this asymmetry indeed arises from the Boltzmann factor. In fact, the slope of the semi-logarithmic plots is in accord with the given temperature T=300K.[4]

---

[4] For this demonstration we use $\eta$=10$^{-5}$cm$^{-1}$. Such a small width parameter is required here, otherwise the lineshape is distorted by the Lorentzian tail associated with thid unphysical width papameter.

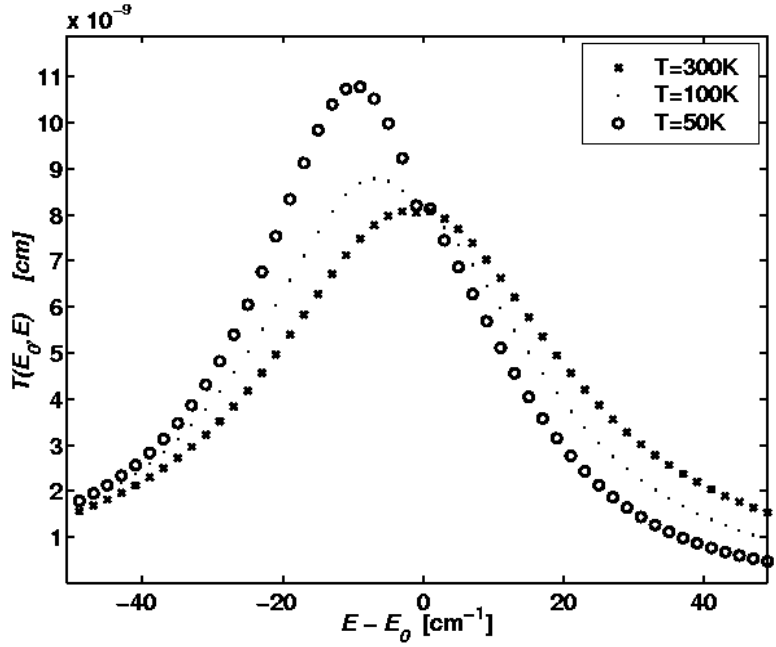

Fig. 4. Same as Fig. 3b for the choice of parameters $\Delta E$=2000cm$^{-1}$ , $\Gamma^{(L)}=\Gamma^{(R)}$=160cm$^{-1}$, $V$=200 cm$^{-1}$ $\kappa$=50 cm$^{-1}$ and $N$=3, shown for three temperatures.

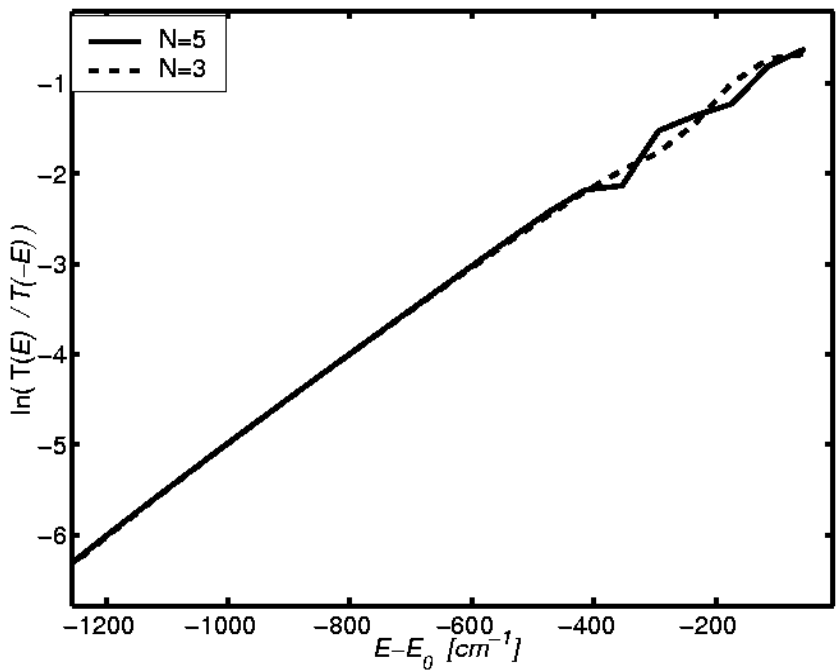

Fig. 5. A semi-logarithmic plot of the asymmetry in the differential transmission, computed for the resonance case $\Delta E$=0cm$^{-1}$. Other parameters are $\Gamma$=160 cm$^{-1}$, $V$=200 cm$^{-1}$, $\kappa$=10 cm$^{-1}$ and T=300K. See text for more details. This calculation is sensitive to numerical artifacts resulting from using finite $\eta$, and $\eta=10^{-5}$ was used here after testing for insensitivity to $\eta$ in this range..

Finally, consider the dependence of the transmission on the thermal coupling strength $\kappa$. Figure 4 shows that the elastic peak decreases with increasing temperature, indicating that the coupling to the thermal environment causes a decrease in the elastic transmission due to the increasing importance of inelastic channels. On the other hand, examining the dependence of the elastic transmission on the thermal coupling strength $\kappa$ reveals a more complex picture. Fig. 6 shows that while at finite temperature the elastic $(E = 0)$ transmission decreases with increasing $\kappa$, at $T=0$ the opposite is true. This last observation is in accord with recent work[32],[51] that shows that in electron-transmission models where coupling to a phonon bath is affected only in the barrier region, such coupling enhances the elastic (and therefore the overall) transmission flux at low temperature, and may be traced to the fact[57] that when a static barrier to transmission becomes amenable to structural relaxation (an attribute of coupling to phonons), the dominating effect at T=0 is lowering the barrier to transmission.

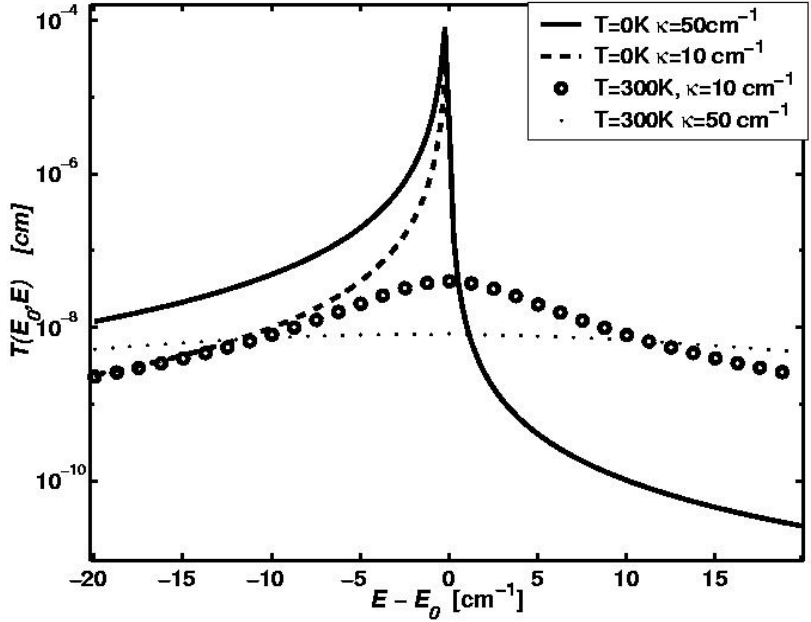

Fig. 6. The energy resolved quasi-elastic transmission peak shown for different temperatures and thermal-coupling strengths. Other parameters are $\Delta E$=2000cm$^{-1}$ ,

$\Gamma$=160 cm$^{-1}$, $V$=200 cm$^{-1}$, N=3. The results are insensitive to the choice of η (here taken $\eta$=0.1cm$^{-1}$).

To conclude this section it should be emphasized again that in the context of our present discussion of electron transmission through molecular bridges the effect of coupling to the thermal environment on the quasi-elastic component of the transmission is not very important because (a) the final energy spectra as seen in Figs (3-6) cannot be resolved in such experiments, and (b) the effect of the correction obtained by the present calculation on the important issue of heat release on the molecular bridge during the transmission is small. In this sense the approximation used in Ref. [44], which accounts for the bulk of the inelastic part of the transmission, is adequate.

## 4. Thermal relaxation effects in the strong coupling limit

The Redfield approximation employed in the previous section is a second-order expansion in the system-thermal bath coupling. Furthermore, this coupling is assumed not to change the equilibrium distribution in the bath. While the latter assumption may be sometimes justified on the basis of size (small system, macroscopic bath) the low order expansion is necessarily a weak coupling approximation. When the coupling is strong one needs to account for distortions in the thermal bath that couple to the electronic transitions. In treatments of some models for electron-phonon coupling in infinite systems this is done within polaron or soliton theories. Here we present such a calculation for our problem of electron transmission through a finite molecular bridge. The importance of local distortion of the nuclear configuration in the process of electron transmission through molecular junctions has been recognized for some time.[28,29,31-33,51,58-64]

Our model is the same as that presented in Sect. 2, however we now limit ourselves to a thermal environment represented by a set of Harmonic oscillators

$$H_B = \sum_\alpha \left( \frac{p_\alpha^2}{2m_\alpha} + \frac{1}{2} m_\alpha \omega_\alpha^2 x_\alpha^2 \right) \tag{38}$$

where $p_\alpha$, $x_\alpha$, $m_\alpha$ and $\omega_\alpha$ are respectively the momenta and coordinates operators, the masses and the frequencies of the harmonic bath modes $\{\alpha\}$. Also, the molecule-bath coupling is now taken to be linear in the phonons coordinates, i.e. Eq. (4) is replaced by the explicit form

$$F = \sum_{n=1}^{N} \sum_{\alpha} \frac{1}{2} c_{\alpha}^{n} x_{\alpha} \mid n >< n \mid \qquad (39)$$

Disregarding the metal electrodes for now we consider the molecule-thermal bath system described by the Hamiltonian

$$H_{MB} = H_M + H_B + F \qquad (40)$$

and apply to it the unitary transformation known as the small polaron transformation[52-54]

$$\tilde{H} = UHU^{-1} \qquad (41)$$

$$U \equiv U_1 U_2, ..., U_{N-1} U_N \qquad (42)$$

where

$$U_n = \exp(-i \mid n >< n \mid \Omega_n) \quad ; \quad \Omega_n = \sum_{\alpha} \Omega_{n,\alpha} \qquad (43)$$

and where

$$\Omega_{n,\alpha} = \frac{c_{\alpha}^{n} p_{\alpha}}{2 m_{\alpha} \omega_{\alpha}^{2}} \qquad (44)$$

The transformed Hamiltonian takes the form[54]

$$\tilde{H} = \sum_{n=1}^{N} E_n \mid n >< n \mid + \tilde{F} + H_B + H_{shift} \qquad (45)$$

where

$$\tilde{F} = V \sum_{n=1}^{N-1} \left( \mid n >< n+1 \mid e^{i(\Omega_{n+1} - \Omega_n)} \quad + \quad \mid n+1 >< n \mid e^{i(\Omega_n - \Omega_{n+1})} \right) \qquad (46)$$

and where

$$H_{shift} = \sum_{n=1}^{N} B_n \mid n >< n \mid \qquad (47)$$

with

$$B_n \equiv -\frac{1}{8} \sum_{\alpha} \frac{(c_{\alpha}^{n})^2}{m_{\alpha} \omega_{\alpha}^{2}} \qquad (48)$$

In an infinite system, where all sites $n$ are equivalent, $B_n$ is the same for all $n$. The uniform shift defined by Eqs. (47) may therefore be disregarded. In the present application the situation is different for two reasons. First, for sites close to the metal interfaces phonon reorganization may be different than for sites in the interior of the molecular bridge. For future reference we shall denote $B_n = B + \delta B_n$ where B is the value of $B_n$ for an interior bridge site far from the bridge-electrode interfaces. Secondly, the (negative) shift $B$ effectively changes the position of the bridge energies relative to the Fermi energies of the metal leads, thus affecting the electron transmission in an essential way.

The shift $B_n$ is a special case of the reorganization energy characterizing shifted Harmonic potential surfaces. Generally, the reorganization energy associated with the electronic transition $n \leftrightarrow n'$ between states characterized by such surfaces is

$$E_R^{(n,n')} = \sum_\alpha \omega_\alpha (\lambda_\alpha^{(n,n')})^2 = \sum_\alpha \frac{(c_\alpha^n - c_\alpha^{n'})^2}{8 m_\alpha \omega_\alpha^2} \tag{49}$$

where

$$(\lambda_\alpha^{(n,n')})^2 = \frac{(c_\alpha^n - c_\alpha^{n'})^2}{8 m_\alpha \omega_\alpha^3} \tag{50}$$

In terms of the spectral density

$$J_{n,n'}(\omega) = \frac{\pi}{2} \sum_\alpha \frac{(c_\alpha^n - c_\alpha^{n'})^2}{m_\alpha \omega_\alpha} \delta(\omega - \omega_\alpha) \tag{51}$$

$E_R^{(n,n')}$ is given by

$$E_R^{(n,n')} = \int_0^\infty \frac{1}{4\pi\omega} J_{n,n'}(\omega) d\omega \tag{52}$$

Next rewrite $\tilde{F}$ in the form $\tilde{F} = <\tilde{F}> + \left(\tilde{F} - <\tilde{F}>\right)$. Including also the coupling to the metal leads, the transformed Hamiltonian takes a form analogous to (7). Assuming again that only levels $|1>$ and $|N>$ of the bridge are coupled to the continuous manifolds that represent the metal leads we get (again $J \equiv (L,R)$ denotes the metal leads)

$$H = \bar{H}_M + H_B + \bar{F} + H_J + \bar{H}_{JM} \tag{53}$$

with

$$\bar{H}_M = \bar{H}_0 + \bar{V} \tag{54}$$

$$\bar{H}_0 = \sum_{n=0}^{N+1} \bar{E}_n |n><n| \quad ; \quad \bar{E}_n = E_n + B_n \tag{55}$$

$$\bar{V} = V \sum_{n=0}^{N} \left( |n><n+1|/<\Theta_{n,n+1}> + |n+1><n|/<\Theta_{n+1,n}> \right) \tag{56}$$

$$H_J = \sum_l E_l \ |l><l| + \sum_r E_r \ |r><r| \tag{57}$$

$$\bar{H}_{JM} = \sum_l \bar{V}_l + \sum_r \bar{V}_r \tag{58}$$

$$\bar{V}_l = V_{l,1}\Theta_{l,1} \mid l><1\mid + V_{1,l}\Theta_{1,l} \mid 1><l\mid$$
$$\bar{V}_r = V_{r,N}\Theta_{r,N} \mid r><N\mid + V_{N,r}\Theta_{N,r} \mid N><r\mid$$
(59)

$$\bar{F} = V\sum_{n=0}^{N}\left\{\mid n><n+1\mid\delta\Theta_{n,n+1} + \mid n+1><n\mid\delta\Theta_{n+1,n}\right\}$$
(60)

where $\mid N+1>$ is the level $\mid f>$,

$$\Theta_{n,n'} = e^{-i(\Omega_n - \Omega_{n'})} \quad ; \quad \delta\Theta_{n,n'} = e^{-i(\Omega_n - \Omega_{n'})} - \left\langle e^{-i(\Omega_n - \Omega_{n'})}\right\rangle$$
(61)

and where, in evaluating Eq. (59) and the corresponding terms in (60) one should keep in mind that, in our model, the left and right manifolds are not associated with phonon shifts, i.e. $\Omega_l = \Omega_r = \Omega_0 = \Omega_f = 0$. The averages $<\Theta_{n,n'}>$ that appear in Eqs. (53)-(61) are over the distribution of the phonon bath, that is assumed here to remain thermal, i.e. $<\Theta_{n,n'}> = Tr_B\left(e^{-\beta H_B}\Theta_{n,n'}\right)\Big/Tr_B\left(e^{-\beta H_B}\right)$.

Note that in distributing the coupling terms between $F$ and $V$ in Eqs. (56), (58)-(60) we have opted not to impose the separation $\Theta = <\Theta> + \delta\Theta$ on the terms appearing in Eqs. (59) so that Eq. (60) does not contain terms (like $\delta\Theta_{1,l}$ and $\delta\Theta_{N,r}$) associated with the coupling of the bridge to the electrodes. It is easy to show that if the continua of $\{l\}$ and $\{r\}$ states are smooth and uniform in an energy range large relative to the reorganization energies associated with the $1\rightarrow\{l\}$ and $N\rightarrow\{r\}$ transitions, respectively, the corresponding self energies are not affected by $\Theta$.[5]

Comparing Eqs. (53)-(60) to (1)-(4), (7) we see a similar structure, except that the tight binding coupling, Eq. (56) is now dressed by thermal terms $<\Theta>$ and the thermal coupling, Eq. (60), connects different electronic states in the local $\{n\}$ representation of the bridge. The important point is that in the transformed Hamiltonian the thermal coupling term $\bar{F}$ is small as long as $V$ is small, and with this assumption the Redfield procedure can be used to find equations of motion for the (reduced) density matrix of the system. We use the reduction procedure described in Sect. 3 and Appendix

---

[5] For example, consider the width $\Gamma_{N,\alpha_N} = 2\pi\sum_r\sum_{\alpha_r}\mid V_{Nr}\mid^2 <\alpha_N\mid\Theta_{Nr}\mid\alpha_r>\mid^2 \delta(E_{N,\alpha_N} - E_{r,\alpha_r})$ of a particular vibronic level $(N,\alpha_N)$ due to its coupling to the vibronic continuum $(r,\alpha_r)$. Here $\alpha_N$ and $\alpha_r$ denote vibrational states on the $N$ and $r$ electronic states, respectively. Since the electronic manifold $\{r\}$ is itself broad, the sum over r in this expression effectively eliminates the $\delta$-function, leaving the sum over final nuclear states $\sum_{\alpha_r} <\alpha_N\mid\Theta_{Nr}\mid\alpha_r>\mid^2 = 1$, irrespective of the details of $\Theta_{Nr}$.

B, whereupon the system consists of the bridge states together with the incoming state $|0>$ of the $\{l\}$ manifold and an outgoing state $|f>$ of the $\{r\}$ manifold. As discussed in Ref. [42], the Redfield procedure should be carried out in the representation that diagonalizes $\bar{H}_M = \bar{H}_0 + \bar{V}$, and we should carry out this diagonalization, find the Redfield equations in this diagonalized representation, then transform back to the original local representation. This leads to[6]

$$\dot{\sigma}_{n,n'} = 0 = -i[\bar{H}_0 + \bar{V}, \sigma]_{n,n'} + \left[ -\frac{1}{2}\Gamma_R(\delta_{n,N} + \delta_{n',N}) - \frac{1}{2}\Gamma_L(\delta_{n,1} + \delta_{n',1}) \right]\sigma_{n,n'}$$
$$+ \sum_m \sum_{m'} R_{n,n',m,m'}\sigma_{m,m'} \quad ; \quad n,n' = 0,1...N, f \quad (n+n' \neq 0)$$

(62)

where $\Gamma_R$ and $\Gamma_L$ are defined as before, Eq. (17),[5] and where $R_{n,n',m,m'}$ are linear combinations of transforms of correlation functions such as[7]

$$\int_0^\infty d\tau \left\langle \delta\Theta_{\nu',\mu}(\tau)\delta\Theta_{\mu,\nu}(0) \right\rangle e^{-i\bar{E}_{\mu,\nu}\cdot\tau} \; ; \; \bar{E}_{\alpha,\beta} = \bar{E}_\alpha - \bar{E}_\beta \,.$$

(63)

The thermal functions $<\Theta>$ and $<\delta\Theta(\tau)\delta\Theta(0)>$ that enter in Eq. (62) are readily calculated for the harmonic bath model used here. Considerable simplification may be achieved by invoking the *local mode approximation*[19,65-67], which relies on the local nature of the electron-phonon coupling in order to assume that different sets of modes are shifted for different electronic states in the local site representation. Under this approximation each site $n$ is associated with a distinct set of modes $\{\alpha n\}$ whose equilibrium positions are shifted when the excess electron is localized on that site. In this case the operators $\Omega_n$ and $\Omega_{n'}$ commute for $n \neq n'$.

Consider first the renormalized coupling elements $\bar{V}$ that contribute to the coherent part of the evolution. Standard calculation yields (Ref. (&&&down)[54], p. 533-550)

---

$$\left\langle \Theta_{n,n'} \right\rangle = \exp\left(-S_{n,n'}\right) \tag{64}$$

$$S_{n,n'} = \frac{1}{2}\sum_{\alpha} (\lambda_{\alpha}^{(n,n')})^2 \ (2\overline{n}_{\alpha}+1) \quad ; \quad \overline{n}_{\alpha} = \left(e^{\omega_{\alpha}/k_BT}-1\right)^{-1} \tag{65}$$

$\left\langle \Theta_{n,n'} \right\rangle$ is the averaged Franck-Condon factor for transitions between electronic states $n$ and $n'$ with thermally equilibrated nuclear populations. In the classical limit $k_BT >> \hbar\omega_{\alpha}$

$$S_{n,n'}^{(cl)} = \sum_{\alpha} (\lambda_{\alpha}^{(n,n')})^2 \ k_BT/\omega_{\alpha} \tag{66}$$

In terms of the spectral density $J_{n,n'}(\omega)$, Eq. (51), we find

$$S_{n,n'} = \frac{1}{8\pi}\int_0^{\infty} \frac{J_{n,n'}(\omega)\coth(\omega/2k_BT)}{\omega^2}d\omega \tag{67}$$

and

$$S_{n,n'}^{(cl)} = \frac{k_BT}{4\pi}\int_0^{\infty} \frac{J_{n,n'}(\omega)}{\omega^3}d\omega \tag{68}$$

The following comments are in order: First, in the local mode approximation where different sets of modes are shifted for different electronic states we have $(c_{\alpha}^n - c_{\alpha}^{n'})^2 = (c_{\alpha}^n)^2 + (c_{\alpha}^{n'})^2$ where, for any given mode $\alpha$, at least one of the terms on the right vanishes. In this case $J_{n,n'}(\omega)$ becomes

$$J_{n,n'}(\omega) = J_n(\omega) + J_{n'}(\omega) \tag{69}$$

$$J_n(\omega) = \frac{\pi}{2}\sum_{\alpha} \frac{(c_{\alpha}^n)^2}{m_{\alpha}\omega_{\alpha}}\delta(\omega-\omega_{\alpha}) \tag{70}$$

$$S_{n,n'} = S_n + S_{n'} \tag{71}$$

$$S_n = \frac{1}{2}\sum_{\alpha} (\lambda_{\alpha}^{(n)})^2 (2\overline{n}_{\alpha}+1) = \int_0^{\infty} d\omega \frac{J_n(\omega)}{8\pi\omega^2}(2\overline{n}(\omega)+1) \quad ; \quad (\lambda_{\alpha}^{(n)})^2 = \frac{(c_{\alpha}^n)^2}{8m_{\alpha}\omega_{\alpha}^3} \tag{72}$$

and similarly

$$E_R^{(n,n')} = E_R^{(n)} + E_R^{(n')} \quad ; \qquad E_R^{(n)} = \sum_{\alpha} (\lambda_{\alpha}^{n})^2\omega_{\alpha} \quad = \int_0^{\infty} \frac{1}{4\pi\omega}J_n(\omega)d\omega \tag{73}$$

Secondly, a standard model for the bath spectral density is[68]

$$J(\omega) \sim \omega^s e^{-\omega/\omega_c} \tag{74}$$

where the cutoff $\omega_c$ corresponds to the upper bound on the phonon frequency. For s<2 $S_{n,n'}$ diverges due to the $\omega \to 0$ divergence of the integrand, therefore $\overline{V}_{n,n'} = 0$ for all $n$ and $n'$ and this contribution to the coherent transport will be completely damped. We expect however that this observation is not relevant to the present case because due to the local character of the electron phonon interaction, the coupling $\left(c_\alpha^n\right)^2$ should go to 0 at least as fast as $\omega$ when $\omega \to 0$. Noting also that the phonon mode density behaves as $\omega^2$ as $\omega \to 0$ we get $s \geq 2$ in Eq. (74).

Even if $s \geq 2$, Eqs. (56), (64)-(65) imply that the local relaxation of the nuclear configuration about each electronic state leads to damping of the direct coupling term $\bar{V}$ and strongly reduces its contribution to the coherent transmission. As seen above, the amount of this damping is strongly sensitive to the low frequency cutoff of $J(\omega)$. In the calculations described below we have used a special case of (74) in the form

$$J(\omega) = 2\pi E_R (\tau_c \omega)^3 e^{-\tau_c \omega} \qquad ; \qquad \tau_c = (\omega_c)^{-1} \tag{75}$$

where the constants in front of $\omega^3 e^{-\tau_c \omega}$ where chosen such that Eq. (73). In the classical limit this choice leads to

$$S_n^{(cl)} = \frac{k_B T}{4\pi} \int_0^\infty \frac{J_n(\omega)}{\omega^3} d\omega = \frac{k_B T E_R}{2\omega_c^2} \tag{76}$$

This special-case result is less important than the general observations: (a) The damping factors $<\Theta>$, which are essentially thermally averaged Franck-Condon factors, diminish strongly when the temperature increases above the characteristic mode energies. (b) In a system in which the electronic transition is strongly coupled to many low frequency modes (e.g in a polar solvent) the $\bar{V}$ term in Eq. (62) may be disregarded for room temperature processes. (c) In the Markovian limit, $\omega_c \to \infty$, $S \to 0$ so $\bar{V}$ retains its bare value $V$. (d) The particular choice of the low frequency cutoff in $J(\omega)$ does not affect correlation functions such as (63) (see below).

To obtain the time evolution according to Eq. (62) we also need to evaluate the correlation functions $C_{k,l,m,n} \equiv \left\langle \delta\Theta_{k,l}(\tau)\delta\Theta_{m,n}(0) \right\rangle$ that enter the rates $R$. Again, such averages can be evaluated using standard harmonic oscillator algebra.[54],[69] Explicit expressions for $R$ and for these correlation functions are given in Appendix C, where we show that in the local mode approximation they be expressed in terms of functions $K_n(t)$ defined for the local bridge states

$$K_n(t) = \exp\left\{ \sum_\alpha \left[ e^{i\omega_\alpha \tau} (\lambda_\alpha^{(n)})^2 \bar{n}_\alpha + e^{-i\omega_\alpha \tau} (\lambda_\alpha^{(n)})^2 (1 + \bar{n}_\alpha) - (\lambda_\alpha^{(n)})^2 (1 + 2\bar{n}_\alpha) \right] \right\}$$

$$= \exp\left( -\int_0^\infty \frac{J_n(\omega)}{4\pi\omega^2} \left[ (2\bar{n}(\omega) + 1) - \bar{n}(\omega)e^{i\omega t} - (\bar{n}(\omega) + 1)e^{-i\omega t} \right] \right) \tag{77}$$

To gain tractable physically significant models it is useful to consider also the correlation function associated with the phonon operators $F_{n,n}$ defined in Eq. (4), which for the model (39) take these form

$$F_{n,n} = \sum_\alpha \frac{1}{2} c_\alpha^n x_\alpha \tag{78}$$

In the local mode approximation $< F_{n,n}(t)F_{n',n'}(0) > = \delta_{n,n'} C_n(t)$ where

$$C_n(t) = \sum_\alpha \frac{(c_\alpha^n)^2}{8 m_\alpha \omega_\alpha} \left[ (\bar{n}_\alpha + 1)e^{-i\omega_\alpha t} + \bar{n}_\alpha e^{i\omega_\alpha t} \right]$$

$$= \frac{1}{4\pi} \int_0^\infty d\omega J_n(\omega) \left[ (\bar{n}(\omega) + 1)e^{-i\omega t} + \bar{n}(\omega)e^{i\omega t} \right] \tag{79}$$

and in the classical limit, $k_B T >> \omega$,

$$C_n(t) = (k_B T / 2\pi) \int_0^\infty d\omega \cos(\omega t) J_n(\omega) / \omega \tag{80}$$

This is essentially a Fourier transform of a function, $J_n(\omega)/\omega$, whose width is of order $\omega_c \equiv \tau_c^{-1}$. Furthermore, using Eq. (73), we get

$$C_n(t = 0) = 2 k_B T E_R^{(n)} \tag{81}$$

For example, to approximately accommodate a model like Eq. (6) with these restrictions we need to take $C(t) = 4 k_B T E_R \tau_c c(t)$, where $c(t) = (1/2)\omega_c e^{-\omega_c|t|}$ becomes $\delta(t)$ in the Markovian limit. Note however the the spectral density $J(\omega)$ associated with this model (given by Eq. (86) below) leads to $S = \infty$ in (72). On the other hand, modifying $J(\omega)$ by simply imposing a low frequency cutoff, $J(\omega) = 0$ for $\omega < \omega_L$ on (86) does not appreciably affect the time dependence of $C(t)$ provided that $\omega_L << \omega_c$.

A link between the models used in the weak thermal coupling limit (Section 3 and Ref. [44]) is obtained from the easily verified relationship[70]

$$K_n(t) = \exp\left[ -it \int_0^\infty \frac{1}{4\pi\omega} J_n(\omega) d\omega \right] \exp\left[ -\int_0^t dt_1 \int_0^{t_1} dt_2 C_n(t_2) \right] =$$

$$= \exp\left[ -it E_R^{(n)} - \int_0^t dt_1 \int_0^{t_1} dt_2 C_n(t_2) \right] \tag{82}$$

Thus, the Markovian weak coupling case ($\tau_c \to 0$ limit of Eq. (6)), $C_n(t) = \kappa_n \delta(t)$ leads to[8]

$$K_n(t) = \exp\left[-iE_R^{(n)}t - (1/2)\kappa_n |t|\right]. \tag{83}$$

and we have already argued that, up to a numerical factor of order 1, $\kappa_n = k_B T E_R^{(n)} \tau_c$. Another model for $K_n(t)$ is obtained by considering the strong coupling limit of Eq. (77).[71] In this limit $K_n(t)$, which vanishes at $t \to \infty$, is assumed vanishingly small already for $t$ short enough to justify expansion of the exponent in (77) to order $t^2$. This yields (using (73))

$$K_n(t) = \exp\left\{-iE_R^{(n)}t - \frac{1}{2}D_n^2 t^2\right\} \tag{84}$$

where

$$D_n^2 = \int_0^\infty d\omega \frac{J_n(\omega)}{4\pi}\left(2\bar{n}(\omega)+1\right)\xrightarrow{k_B T \gg \omega} 2k_B T E_R^{(n)} \tag{85}$$

The forms (83) and (84) are obtained as limiting cases of a model[70] that uses a spectral density of the Debye form

$$J_n(\omega) = 8E_R^{(n)}\frac{\tau_c \omega}{1+(\tau_c \omega)^2}\Theta(\omega) \tag{86}$$

In the classical limit (80) of Eq. (79) it yields[70][9]

$$C_n(t) = 2E_R^{(n)}k_B T e^{-t/\tau_c} \tag{87}$$

Using this in (82) leads to (83) with $\kappa_n = k_B T E_R^{(n)}\tau_c$ in the limit $(2E_R k_B T)^{1/2}\tau_c \ll 1$, and to (84) with $D_n^2 = 2k_B T E_R^{(n)}$ in the opposite limit $(2E_r k_B T)^{1/2}\tau_c \gg 1$. Again the model (86) implies $S_n \to \infty$. In the spirit of the discussion below Eq. (81) it is easy to show that the forms (83) and (84) for $K_n$ remain intact in the Markovian $\left(\omega_c \to \infty\right)$ and the strong

---

[8] The appearance of |t| in (83) results from the easily proven identity

$$K_n(-t) = \exp\left[itE_R^{(n)} - \int_0^{|t|}dt_1\int_0^{t_1}dt_2 C_n(-t_2)\right].$$

[9] In fact (86) with the high T limit of (79) yields $C_n(t) = e^{-t/\tau_c}(2E_R^{(n)}k_B T - iE_R^{(n)}\tau_c^{-1})$, where the second term was neglected in (87). Keeping this term and using (82) leads to (83) without the $E_R^{(n)}$ term in the exponent. Again, this makes little difference in the high $T$ limit.

coupling limits, respectively, if a small low frequency cutoff is imposed on the Ohmic model

Having found explicit expressions or computational procedures for the elements of $\bar{V}$ and $R$ in Eq. (62) we may proceed to solve the set of kinetic equation (62) for the transmitted current, as described in Sect. 4. The energy resolved transmission is again given by $\lim_{\eta \to 0} \eta \sigma_{ff}$ and the total transmission is obtained by summing this result over all final states in the $\{r\}$ manifold.

The discussion of Fig. 6 above has referred to the effect of coupling to phonons on the elastic component of the electron transmission flux. Next we use the computational scheme outlined above to study the effect of this coupling on the overall transmission. Figures 7 and 8 display the transmission flux out of a particular energy level $E_0$ in the left electrode, plotted against the reorganization energy $E_R$ that represents the strength of coupling to environmental modes. Recalling the form of the electron-phonon coupling in our model, Eqs. (4) and (39), there are two principal ways in which this coupling affects the electron transmission. First, this coupling causes a relative vertical shift of the parabolic potential surfaces associated with the different electronic states, (cf Eqs. (47) and (48)). Secondly, it leads to a relative horizontal shift of these surfaces, effectively decreasing the interstate coupling by the corresponding Franck Condon (FC) factors; $|\bar{V}| < |V|$ in, e.g., Eq. (56). The combination of both effects leads to the appearance of $E_R$ in the activation factor,

$$FC \sim \exp\left(-(E_{DA} - E_R)^2 / 4k_B T E_R\right) \tag{88}$$

in the semiclassical electron transfer rate.[72,73] Qualitatively these two effects can be designated as a renormalization of the potential surfaces and as phonon-induced friction, respectively. It is useful to study these effects separately as is often done in theoretical studies of friction effects on chemical reactions. This is done in Figure 7, where the dashed and full lines correspond to the flux obtained from the calculation described above, while the dotted and dashed-dotted lines are obtained from a similar calculation using a model in which $H_{shift}$ of Eq. (47) is set to zero. Here and in Fig. 8 the factor $S$, Eq. (72), was computed using the spectral density (75). The dashed and dotted lines were calculated using $\omega_c = 1000 \text{cm}^{-1}$ that correspond to the Markovian limit, Eq. (83), while the full and dashed-dotted lines where computed in the strong coupling limit, Eq. (84)-(85).

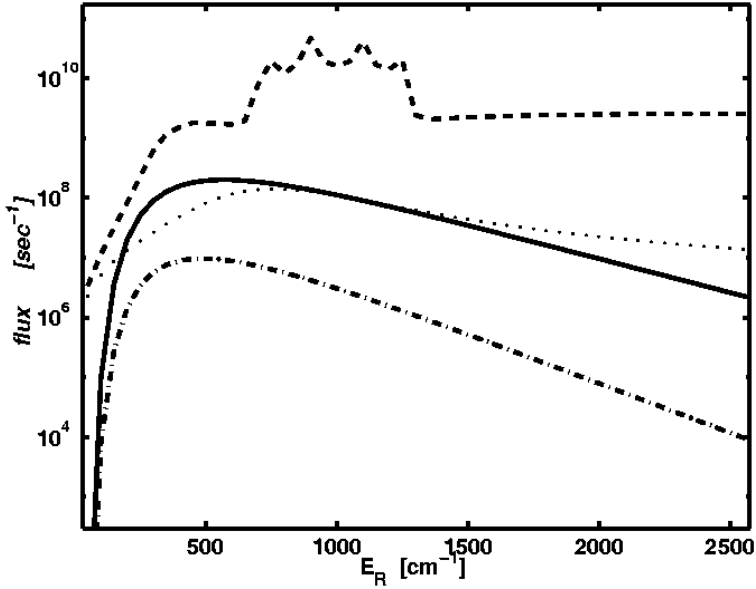

Fig. 7. The transmitted flux plotted against the electron-phonon interaction strength expressed by the reorganization energy $E_R$, for a system with $N=4$ bridge levels characterized by $\Gamma^L=\Gamma^R=160$ cm$^{-1}$, $V=200$ cm$^{-1}$, $\Delta E=1000$cm$^{-1}$ and $T=300$K. The dashed and dotted lines were computed using $\omega_c=1000$cm$^{-1}$ in the Markovian limit, Eq. (83). Note that for large $E_R$ the condition for validity of the Markovian limit, $(2E_R k_B T)^{1/2} \tau_c \ll 1$, may not hold. The full and dashed-dotted lines were computed with $\omega_c=10$cm$^{-1}$, using the strong coupling expression (84). For the calculation that yields the dotted and dashed-dotted lines the vertical shift associaed with the reorganization, $H_{shift}$ (Eq. (47), was set to zero.

Consider first the dashed line of Fig. 7. When the bare energy gap $\Delta$E is small enough, the phonon induced vertical shift may bridge the lowest bridge levels into resonance with the incoming energy $E_0$, leading to the observed resonance structure. This structure corresponds to the four coupled electronic sites of the molecular bridge used in this calculation. Because the vibrational spectrum (e.g., Eq. (75)) used in our model is dense and relatively smooth, no further structure is seen, and the lack of dependence of the dashed line on $E_R$ for $E_R \geq 1300 cm^{-1}$ reflects the fact that in this range the initial energy faces this smooth continuum of phonon states. When the vertical shift is absent (dotted line) so that the effective energy gap retains its bare value $\Delta$E, the rise and subsequent fall of the flux with increasing $E_R$ reflect both the effect of the nuclear degrees of freedom on the effective activation energy and their role as a dissipative medium, that in the semiclassical ($k_B T \gg \hbar\omega_c$) limit combine to yield the thermally averaged Franck-Condon factor, Eq. (88). The parameters used in the other two (full and dash-dotted) lines correspond to this limit. Overall, it is seen that, as in the standard theory of electron transfer coupling to the environment can either enhance or inhibit the electron flux. At this point it is perhaps worthwhile to emphasize the obvious fact that unlike in the standard molecular electron transfer processes, electron flux between two electrodes can take place without coupling to the nuclear environment.

## 5. Discussion and conclusions

This paper has developed a framework for evaluating electron transmission through models of molecular bridges in the presence of thermal interactions. We have considered two approaches, both based on the Redfield approximation. The first assumes that the thermal coupling is weak and employs the Redfield procedure on the given Hamiltonian in which the coupling between the system and its thermal environment is represented by terms of the form (4) with the thermal bath operators $F_{n,n}$ satisfying Eq. (6). The second starts with a similar model, specialized to the case of a harmonic thermal bath and $F$ linear in the phonon coordinates and applies the small polaron transformation leading to a set of equations (54)-(61) similar to that used in the first approach but with renormalized coupling coefficients. In particular the tight binding coupling $V$ in Eq. (1) is replaced by the renormalized coupling $\overline{V}$, Eq.(56) and the thermal coupling operator (4) is replaced by Eq. (60). An important, and sometimes doubtful ingredient in both approaches is the assumption that the thermal bath remains in Boltzmann equilibrium during the steady state electron flow through the system. In the first approach this assumption is a reasonable consequence of the assumed weak system-thermal bath coupling. In the second approach that allows strong thermal coupling this approximation has to rely on another assumption, that nuclear thermal relaxation at each electronic state is fast on the timescale of any electronic transition. The latter assumption is often valid in the hopping regime, where the bridge electronic levels are physically populated and transport proceeds predominantly by electronic transitions between sites with thermally relaxed nuclear distributions. It is however questionable in situations where the electron injection energy is not in resonance with the bridge levels.

Obviously, the equivalence of the two schemes for the case of a harmonic thermal bath coupled linearly to the electronic transition implies that the dephasing process associated with the term (4) of the Hamiltonian (7) in the first approach is related to the reorganization energy (49). This equivalence can be established quantitatively using Eq. (82) as discussed in Sect. 4. On the qualitative level it is of interest to see to what extent the two computational schemes can reproduce the transition from coherent transport to incoherent hopping as the bridge length increases (Fig. 6 and References [42],[56]). Figure 8 show the transmission flux plotted against the number of bridge sites $N$ for different values of the thermal coupling $\kappa$ (and the corresponding reorganization energy $E_R$, assuming the relation $\kappa = 4k_B T E_R \tau_c$, same for all bridge sites). It is seen that both computational schemes lead to the same qualitative dependence on $N$, however they quantitatively differ in the hopping regime by a factor of up to an order of magnitude for large thermal couplings. It should be kept in mind that, while it is usually assumed that the hopping process is characterized by full thermal relaxation in each local site (a picture adopted in the polaronic model of Sect. 4), the validity of this assumption is not a-priori obvious (see below). The opposite limit, where the mean free path of the electron motion is larger than the distance between two bridge sites, is better described by the weak coupling model of Sect. 3. A more advanced theory (see below) is needed to bridge between these limits.

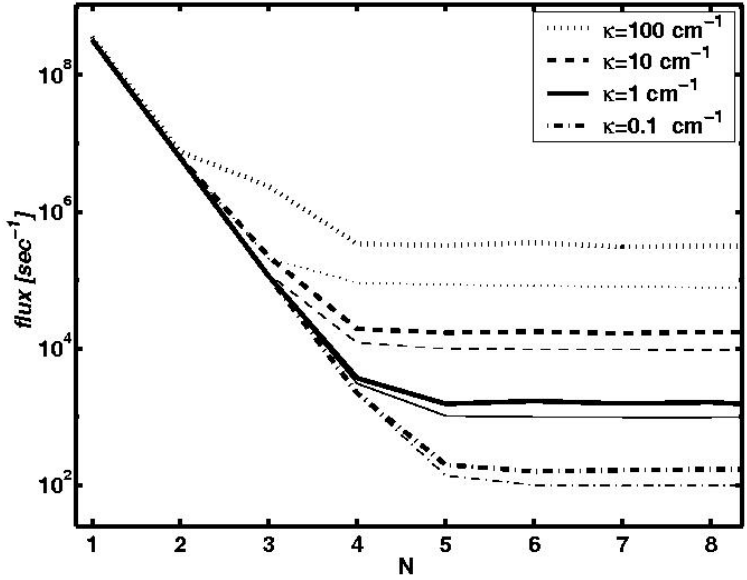

Fig. 8. The transmitted flux vs. bridge length $N$ for different values of the thermal coupling $\kappa$. The other system parameters are $\Gamma^{(L)} = \Gamma^{(R)} = 160$ cm$^{-1}$, $V = 200$ cm$^{-1}$, $\Delta E = 1500$cm$^{-1}$, $\omega_c = 800$ cm$^{-1}$ and $T = 300$K. The thin lines represent results obtained from the weak coupling approximation of Sect. 3. The heavy lines were obtained using the small polaron transformation of Sect. 4 in the Markovian limit.

While the formulation of Sect. 4 obviously reduces to the weak-coupling limit of Sect. 3 in the proper limit, the intermediate case is not properly described by this theory that assumes that the thermal bath is in equilibrium with the instantaneous local electronic state. A full theory of the transition between these two limits should take into account the timescale associated with this relaxation.[19] To see how this timescale may be taken into account consider for simplicity a model with one bridge state connecting the two continuous manifolds that represent the electrodes. The transmission amplitude $\mathcal{T}$ in this case is proportional to $(\Delta\bar{E} - i\Gamma/2)^{-1}$ where $\Gamma$ is the total width of the bridge level due to its interaction with the continua and $\Delta\bar{E} = \Delta E - E_R^{eff}(\tau)$ is the relaxed energy gap between the bridge state and the incoming energy, modified by the effective reorganization energy $E_R^{eff}(\tau)$. (As before $\Delta E = E_1 - E_0$ is the bare energy gap). The transmission time associated with competing relaxation processes in the bridge is[74]

$$\tau = |\mathcal{T}|^{-1}|\partial\mathcal{T}/\partial\Delta E| = \left[\left(\Delta E - E_R^{eff}(\tau)\right)^2 + (\Gamma/2)^2\right]^{-1/2} \qquad (89)$$

A typical timescale of the thermal environment was set by $\omega_c$. Using the ansatz for the time evolution of the effective reorganization energy

$$E_R^{eff}(\tau) = E_R \left[ 1 - e^{-\omega_c \tau} \right] \tag{90}$$

where $E_R$ is, as before, the reorganization energy associated with the fully relaxed intermediate state, gives an equation for $\tau$ that, once solved, gives via (90) the value of the effective reorganization energy $E_R^{eff}(\tau)$ associated with the finite traversal time $\tau$ of the electron through the intermediate level. As an example, Fig. 9 shows $E_R^{eff}(\tau)$ as a function of $\Delta E$ for a model with full reorganization energy $E_R=200\text{cm}^{-1}$ and for two values of $\omega_c$. As is intuitively clear, we see that the effective reorganization energy rapidly decreases with increasing $\Delta E$, so that for bare energy gaps exceeding 0.25eV, say, we quickly approach the weak coupling limit.[10] This physics is missing in the theory presented in Sect. 4, and will be addressed in future work.

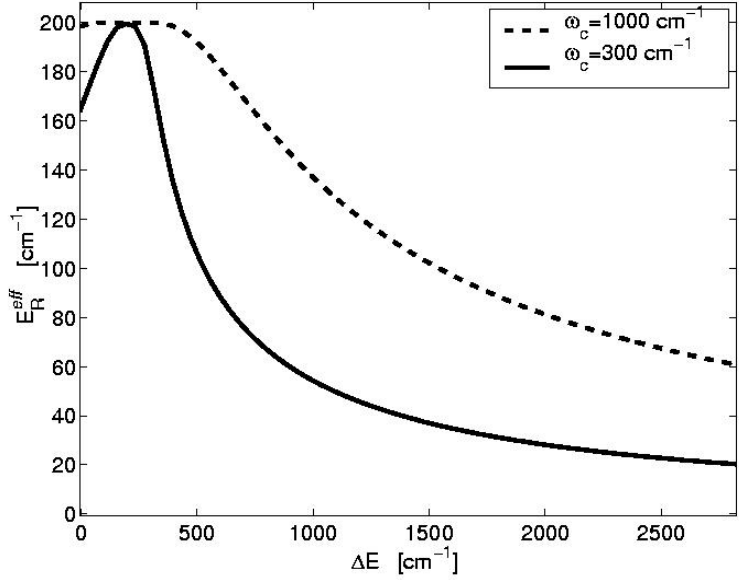

Fig. 9. The effective reorganization energy computed from Eqs. (89)-(90) for a model with one bridge level, with $\Gamma = \Gamma^L + \Gamma^R = 50\text{cm}^{-1}$, T=300K and $E_R=200\text{cm}^{-1}$.

The existence of an intermediate regime, where electron transmission is dominated by thermally induced propagation through the bridge but classical hopping that relies on full local thermal relaxation (i.e. local reorganization) at each bridge site is still not valid, may have important consequences. For example, it may be the reason for the difficulty to account quantitatively for the experimental results of Ref. [56] by purely kinetic schemes, as discussed by Bixon and Jortner in this issue. [75] Notwithstanding possible other factors, e.g. configurational changes following electron (or hole) injection into the bridge, the possibility that the bridge propagation is affected by another physical parameter, essentially the mean free path of the charge carrier, cannot be ruled out at this point.

---

[10] The fact that $E_R^{eff}$ goes through a maximum in Fig. 10 results from the fact $\tau$, the solution to Eqs. (89)-(90) goes trough a maximum in the vicinity of $\tau=\Gamma^{-1}$ as a function of $\Delta E$.

Of lesser conceptual significance, but very important for actual calculation is that the phonon spectral density employed in this work corresponds to intermolecular nuclear motions and should be substantially modified to include high frequency vibrational modes. Such modes cannot be treated semiclassically, and, being strongly underdamped, cannot be assumed to be in thermal equilibrium throughout the transmission process. Including such modes specifically in the calculation requires explicit consideration of the vibronic levels involved as was done in Ref. [42].

Finally, it should be kept in mind that our nearest-neighbor coupling model has to be modified when realistic chain configurations are considered, since exclusive 'through bond' transfer does not necessarily dominates the electron transfer process.[76]

## Appendix A

Here we consider two methods to describe resonance transmission between two continuum manifolds $\{l\}$ and $\{r\}$ through a bridge, represented for simplicity by a single level $|1>$. Method A, which focuses on the incoming state as the driving force that keeps the system in a non-equilibrium steady state, is identical to the route taken by us before.[44] Method B treats the incoming and outgoing states symmetrically as shown below.

*Method A.* A steady state driven by an incoming state $|0>$ is described by the set of equations equivalent to (8)-(13)

$$\dot{c}_0 = -iE_0 c_0 \tag{91}$$

$$\dot{c}_1 = -iE_1 c_1 - iV_{1,0} c_0 - i\sum_{l \neq 0} V_{1,l} c_l - i\sum_r V_{1,r} c_r \tag{92}$$

$$\dot{c}_l = -iE_l c_l - iV_{l,1} c_1 - (\eta/2)c_l \tag{93}$$

$$\dot{c}_r = -iE_r c_r - iV_{r,1} c_1 - (\eta/2)c_r \tag{94}$$

At steady state all coefficients satisfy $c_k(t) = C_k e^{-iE_0 t}$ ; $\qquad k = 0,1,l,r$ with

$$0 = -i(E_1 - E_0)C_1 - iV_{1,0}C_0 - i\sum_{l \neq 0} V_{1,l}C_l - i\sum_r V_{1,r}C_r \tag{95}$$

$$0 = -i(E_j - E_0)C_j - iV_{j,1}C_1 - (\eta/2)C_j \quad ; \quad j = l,r \tag{96}$$

Solving (96) for $C_l$ and $C_r$ in terms of $C_1$ and inserting the results in (95) leads to

$$C_1 = \frac{V_{1,0}C_0}{E_0 - \tilde{E}_1 + (i/2)\Gamma_1(E_0)} \tag{97}$$

where $\Gamma_1(E_0) = \Gamma_1^{(L)}(E_0) + \Gamma_1^{(R)}(E_0)$ and $\tilde{E}_1 = E_1 + \Lambda_1^{(L)}(E_0) + \Lambda_1^{(R)}(E_0)$ with $\Gamma$ and $\Lambda$ defined from $\sum_r |V_{r,1}|^2 /(E_0 - E_r + i\eta/2) = \Lambda_1^{(R)}(E_0) - (1/2)i\Gamma_1^{(R)}(E_0)$ (and similarly for L). Using Eq. (97) in (96) and taking the limit $\eta \to 0$ yields

$$\frac{\eta|C_r|^2}{|C_0|^2} = 2\pi|V_{r,1}|^2 \, \delta(E_r - E_0) \quad \frac{|V_{1,0}|^2}{\left(\tilde{E}_1 - E_0\right)^2 + \left(\Gamma_1(E_0)/2\right)^2} \tag{98}$$

which leads[44] to the (differential) transmission coefficient

$$\mathcal{T}'(E_0,E) = \frac{\Gamma_1^{(L)}(E_0)\Gamma_1^{(R)}(E_0)}{\left(\tilde{E}_1 - E_0\right)^2 + \left(\Gamma_1(E_0)/2\right)^2} \delta(E - E_0) \tag{99}$$

*Method B.* In this alternative approach we handle the incoming and outgoing states in a more symmetrical fashion. To this end we consider again Eq. (95) written in the form

$$0 = -i(E_1 - E_0)C_1 - iV_{1,0}C_0 - iV_{1,f}C_f - i\sum_{l \neq 0} V_{1,l}C_l - i\sum_{r \neq f} V_{1,r}C_r \tag{100}$$

where $|f\rangle$ is one particular final state in the $\{r\}$ continuum. Again using solutions of (96) in (100) we get

$$0 = -i(\tilde{E}_1 - E_0)C_1 - iV_{1,0}C_0 - iV_{1,f}C_f - (1/2)\Gamma_1(E_0)C_1 \tag{101}$$

It is important to keep in mind that because the states 0 and $f$ are unbounded continuous states, the coefficients $C_0$ and $C_f$ scale like $\Omega^{-1/2}$ where $\Omega$ is the normalization volume ($\Omega \to \infty$ should be taken at the end of the calculation), therefore $\Gamma_1$ and $\Lambda_1$ remain as before (these quantities remain finite when $\Omega \to \infty$ because they contain products such $|V_{1r}|^2 \rho_R$ with $\rho_R$ being the density of states in the R manifold that scales like $\Omega$). Eq. (101) together with Eq. (96) written once for $j=0$ and once for $j=f$ constitute 3 coupled linear equations for $C_0$, $C_1$ and $C_f$ that may be solved to yield

$$\left[1 + \frac{|V_{1,f}|^2}{\left(\tilde{E}_1 - E_0 - i/2\Gamma_1(E_0)\right)\left(E_0 - E_f + i\eta/2\right)}\right] C_f = \frac{V_{f,1}V_{1,0}C_0}{\left(E_0 - E_f + i\eta/2\right)} \frac{1}{\left(E_0 - \tilde{E}_1 + i/2\Gamma_1(E_0)\right)}$$

(102)

The second term in the brackets on the l.h.s vanishes in the limit $\Omega \to \infty$. The remaining terms lead to a result of the form (98) (with $f$ replacing $r$) in this limit. We conclude that in the thermodynamic limit the two methods lead to identical results.

## Appendix B

To reduce the set (32)-(37) to a set of steady state equations for the system's density matrix $\sigma$ in the Redfield approximation[36,46,47] we follow the procedure of Ref. [44]. The 'system' consists of the set of bridge states $|1>,...|N>$ (here this set consists only of a single level $|1>$) together with the initial and final continuum states $|0>$ and $|f>$. The Redfield expansion needs to be carried in the representation that diagonalizes the effective system's Hamiltonian that includes now, in addition to the bridge state $|1>$ also the incident state $|0>$ and the final state $|f>$

$$H_0^{eff} = \begin{pmatrix} E_0 & 0 & 0 \\ V_{1,0} & E_1 - i\Gamma_1/2 & V_{1,f} \\ 0 & V_{f,1} & E_f \end{pmatrix}$$

(103)

The procedure therefore starts with a transformation to the representation which diagonalizes this Hamiltonian, following the Redfield procedure in this representation then transforming back to the representation defined in terms of states $|0>$ $|1>$ and $|f>$. It yields[42,44]

$$\dot{\sigma}_{n,n'} = 0 = -i[H_0 + V, \sigma]_{n,n'}^{'} - \frac{1}{2}\Gamma_1\left(\delta_{n,1} + \delta_{n',1}\right)\sigma_{n,n'} + \sum_{n_1}\sum_{n_2} R_{n,n',n_1,n_2}\sigma_{n_1,n_2} \; ; n,n' = 0,1,f$$

(104)

where the prime on the commutator denotes that it has been modified by eliminating the terms incompatible with a steady-state driven by state $|0>$ as discussed under Eq. (25) (see also Ref. (&&&down)[44]), and where the tetradic elements $R_{n_1,n_2,n_3,n_4}$ are be expressed in terms of the correlation function $<F_{1,1}(t)F_{1,1}(0)>$. Also, the sums in (104) are over $n_1, n_2 = 0,1,f$. Solving (104) for the steady state defined by a fixed $\rho_{0,0}$ finally yields the steady-state values of $\sigma_{0,1}$, $\sigma_{1,0}$, $\sigma_{1,1}$, $\sigma_{0,f}$, $\sigma_{f,0}$, $\sigma_{1,f}$, $\sigma_{f,1}$, $\sigma_{f,f}$.

## Appendix C

As discussed in Section 4, the reduced density matrix evolves according to Eq (62). The tensor $R$ is obtained by (numerically) transforming the tensor $\bar{R}$

$$
\bar{R}_{\nu,\nu',\mu,\mu'} = \int_0^\infty \left\langle K_{\mu',\nu'}(0) K_{\nu,\mu}(\tau) \right\rangle e^{-i\bar{E}_{\mu,\nu'}\cdot\tau} d\tau + \int_0^\infty \left\langle K_{\mu',\nu'}(\tau) K_{\nu,\mu}(0) \right\rangle e^{-i\bar{E}_{\nu,\mu'}\cdot\tau} d\tau
$$
$$
-\delta_{\mu',\nu'} \sum_\lambda \int_0^\infty \left\langle K_{\nu,\lambda}(\tau) K_{\lambda,\mu}(0) \right\rangle e^{-i\bar{E}_{\lambda,\mu'}\cdot\tau} d\tau - \delta_{\nu,\mu} \sum_\lambda \int_0^\infty \left\langle K_{\mu',\lambda}(0) K_{\lambda,\nu'}(\tau) \right\rangle e^{-i\bar{E}_{\mu,\lambda}\tau} d\tau \tag{105}
$$

(where $K_{\nu,\mu}(t)$ (in the basis that diagonalizes the bridge Hamiltonian) is a linear combination of terms $V_{n,m}\delta\Theta_{n,m}$ (in the basis of local site states) in line with the procedure discussed above Eq. (62). In Eq. (105), $\bar{E}_{i,j}$ are differences between eigenvalues of the same bridge Hamiltonian. The operators $\delta\Theta$ are defined by Eq. (61). The coupling terms $V_{ij}$ that appear in (105) result from the diagonalization procedure described above Eq. (62). It leads to the fact that the tensor $R$ couples sites that are not nearest neighbors. The correlation functions that appear in (105) have the form

$$
C_{k,l,m,n} \equiv \left\langle \delta\Theta_{k,l}(\tau) \delta\Theta_{m,n}(0) \right\rangle = \left\langle e^{i(\Omega_k(\tau)-\Omega_l(\tau))} e^{i(\Omega_m(0)-\Omega_n(0))} \right\rangle - \left\langle e^{i(\Omega_k-\Omega_l)} \right\rangle \left\langle e^{i(\Omega_m-\Omega_n)} \right\rangle \tag{106}
$$

Explicit expressions for these functions may be found by using standard harmonic oscillator operator algebra.[54] We get

$$
\left\langle e^{i(\Omega_k(\tau)-\Omega_l(\tau))} e^{i(\Omega_m(0)-\Omega_n(0))} \right\rangle =
$$
$$
\exp\left\{ \sum_\alpha \left[ e^{i\omega_\alpha\tau} \lambda_\alpha^{(k,l)} \lambda_\alpha^{(n,m)} \bar{n}_\alpha + e^{-i\omega_\alpha\tau} \lambda_\alpha^{(k,l)} \lambda_\alpha^{(n,m)} (1+\bar{n}_\alpha) - \frac{1}{2}\left[ (\lambda_\alpha^{(k,l)})^2 + (\lambda_\alpha^{(n,m)})^2 \right](1+2\bar{n}_\alpha) \right] \right\} \tag{107}
$$

and

$$
\left\langle e^{i(\Omega_k-\Omega_l)} \right\rangle \left\langle e^{i(\Omega_m-\Omega_n)} \right\rangle = \exp\left\{ -\frac{1}{2} \sum_\alpha \left[ (\lambda_\alpha^{(k,l)})^2 + (\lambda_\alpha^{(n,m)})^2 \right](2\bar{n}_\alpha+1) \right\} \tag{108}
$$

Where

$$
(\lambda_\alpha^{(n,m)})^2 = \frac{(c_\alpha^n - c_\alpha^m)^2}{8 m_\alpha \omega_\alpha^3} \tag{109}
$$

These results can be simplified by invoking the local mode approximation introduced in Sect 4. Eq. (109) then becomes

$$(\lambda_\alpha^{(n,m)})^2 = (\lambda_\alpha^{(n)})^2 + (\lambda_\alpha^{(m)})^2 \tag{110}$$

with

$$(\lambda_\alpha^{(n)})^2 = \frac{(c_\alpha^n)^2}{8m_\alpha\omega_\alpha^3} \tag{111}$$

Note that under this approximation $c_\alpha^n$ and $c_\alpha^m$ cannot be both non-zero for the same $\alpha$ and $m \neq n$, therefore one of the terms on the r.h.s. of Eq. (110) is zero. Eqs. (107) and (108) now take the simpler forms

$$\left\langle e^{i(\Omega_k(\tau)-\Omega_l(\tau))}e^{i(\Omega_m(0)-\Omega_n(0))}\right\rangle = \begin{cases} \exp\left(-S_k - S_l - S_m - S_n\right) & ; & k \neq n \neq l \neq m \\ K_l(\tau)K_k(\tau) & ; & k = n, l = m \neq k \\ K_k(\tau)\exp\left(-S_l - S_m\right) & ; & k = n, l \neq m \neq k \\ K_l(\tau)\exp\left(-S_k - S_n\right) & ; & l = m, k \neq n \neq l \\ \exp\left(-S_m - S_n\right) & ; & k = l, n \neq m \\ \exp\left(-S_k - S_l\right) & ; & k \neq l, n = m \\ \bar{K}_m(\tau)\bar{K}_l(\tau) & ; & k = m, l = n \neq k \\ \bar{K}_m(\tau)\exp\left(-S_n - S_l\right) & ; & k = m, l \neq n \neq m \\ \bar{K}_l(\tau)\exp\left(-S_m - S_k\right) & ; & l = n, m \neq k \neq l \end{cases} \tag{112}$$

and

$$\left\langle e^{i(\Omega_k - \Omega_l)}\right\rangle = \begin{cases} \exp\left(-S_k - S_l\right) & ; & k \neq l \\ 1 & ; & k = l \end{cases} \tag{113}$$

Where $S_n$ is given by Eq. (72) and where

$$K_n(t) = \exp\left\{\sum_\alpha\left[e^{i\omega_\alpha\tau}(\lambda_\alpha^{(n)})^2\bar{n}_\alpha + e^{-i\omega_\alpha\tau}(\lambda_\alpha^{(n)})^2(1+\bar{n}_\alpha) - (\lambda_\alpha^{(n)})^2(1+2\bar{n}_\alpha)\right]\right\}$$
$$\bar{K}_n(t) = \exp\left\{\sum_\alpha\left[-e^{i\omega_\alpha\tau}(\lambda_\alpha^{(n)})^2\bar{n}_\alpha - e^{-i\omega_\alpha\tau}(\lambda_\alpha^{(n)})^2(1+\bar{n}_\alpha) - (\lambda_\alpha^{(n)})^2(1+2\bar{n}_\alpha)\right]\right\} \tag{114}$$

The correlation function (106) then takes the form

$$C_{k,l,m,n} = \begin{cases} 0 & k \neq n \neq l \neq m \\ K_k(\tau)K_l(\tau) - \exp\left(-2\left(S_k + S_l\right)\right) & k = n, l = m \neq k \\ K_k(\tau)\exp\left(-S_l - S_m\right) - \exp\left(-2S_k - S_l - S_m\right) & k = n, l \neq m \\ K_l(\tau)\,\exp\left(-S_k - S_n\right) - \exp\left(-S_k - 2S_l - S_n\right) & l = m, k \neq n \\ 0 & k = l, n \neq m \\ 0 & k \neq l, n = m \\ \bar{K}_m(\tau)\bar{K}_l(\tau) - \exp\left(-2\left(S_m + S_l\right)\right) & ; \quad k = m, l = n \neq k \\ \bar{K}_m(\tau)\exp\left(-S_n - S_l\right) - \exp\left(-2S_m - S_n - S_l\right) & ; \quad k = m, l \neq n \neq m \\ \bar{K}_l(\tau)\exp\left(-S_m - S_k\right) - \exp\left(-2S_l - S_k - S_m\right) & ; \quad l = n, m \neq k \neq l \end{cases} \qquad (115)$$

(In our numerical calculation we neglect the correlation functions associated with the last three lines of (115), since they are smaller than the other nonzero terms). Finally, in a model where the coupling coefficient $\lambda_\alpha^{(n)}$ (and consequently $S_n$ do not depend on the electronic site $n$, the non-zero terms of (115) take the very simple forms $C_{k,l,m,n} = K(\tau)^2 - \exp(-4S)$ for $k = n, l = m$ and $C_{k,l,m,n} = K(\tau)\,\exp(-2S) - \exp(-4S)$ for $k = n, l \neq m$ or $k \neq n, l = m$.

**Acknowledgements**   This research was supported in part by the United States-Israel Bi-national Science Foundation. The work of DS is supported by a fellowship from the Clore Foundation. AN thanks the Institute of Theoretical Physics at UCSB for hospitality during the final stages of this work.

---

[1] While our language refer to electron transport and electron tunneling, hole transport
    and numclear excitation via transient positive ion formation are equally possible.